\documentclass[twocolumn,showpacs,superscriptaddress,prl]{revtex4}

\usepackage{graphicx}
\usepackage{amsmath}
\usepackage{amsfonts}
\usepackage{bm}

\begin{document}

\preprint{}

\title[Josephson Effect with Internal Structures]%
{Josephson Effect between Condensates with Different Internal Structures}

\author{Munehiro Nishida}
 \affiliation{%
 Low Temperature Laboratory, Helsinki University of
 Technology, P.O. Box 2200, FIN-02015 HUT, Finland
 }
 \affiliation{%
 Department of Physics, Waseda University, Okubo, Shinjuku-ku,
 Tokyo 169-8555, Japan
 }
% \email{nishida@kh.phys.waseda.ac.jp}
% \homepage{http://www.phys.waseda.ac.jp/kh/kh/nishida/nishida-e.html}
\author{Noriyuki Hatakenaka}
 \affiliation{%
 Low Temperature Laboratory, Helsinki University of
 Technology, P.O. Box 2200, FIN-02015 HUT, Finland
 }
 \affiliation{%
 NTT Basic Research Laboratories, Atsugi, Kanagawa
 243-0198, Japan
 }
\author{Susumu Kurihara}
 \affiliation{%
 Department of Physics, Waseda University, Okubo, Shinjuku-ku,
 Tokyo 169-8555, Japan
 }

\date{\today}

\begin{abstract}
 A general formula for Josephson current in a wide class of hybrid
 junctions between different internal structures is derived on the basis
 of the Andreev picture.
 The formula extends existing formulae and also enables us to analyze
 novel \textit{B}-phase/\textit{A}-phase/\textit{B}-phase (\textit{BAB})
 junctions in superfluid $^3$He systems, which are accessible to
 experiments.
 It is predicted that \textit{BAB} junctions will exhibit two types of
 current-phase relations associated with different internal symmetries.
 A ``pseudo-magnetic interface effect'' inherent in the system is also
 revealed.
\end{abstract}


\maketitle

%\tableofcontents
Superfluid systems with internal degrees of freedom produce diverse
ordered structures, which provide a new arena for exploring the fertile
physics behind them.
The Josephson effect \cite{josephson65} extracts the global phase
resulting from the spontaneous breakdown of gauge symmetry.
Moreover, it also provides us with information on the internal
structures formed from other broken symmetries since quasiparticles
acquire an excess phase from the internal structures while traveling
through the junction.
In fact, so-called $\pi$ junctions associated with high-$T_c$
superconductors provide convincing evidence for $d$-wave symmetry
\cite{sigrist92,wollman93}, and recent experiments on
Pb-Sr$_2$RuO$_4$-Pb Josephson junctions \cite{jin99} could be explained
by $p$-wave symmetry in Sr$_2$RuO$_4$ \cite{yamashiro98,honerkamp98}.
Moreover, metastable $\pi$ states observed in superfluid $^3$He weak
links \cite{packard98,packard99,avenel00} are considered to be a
signature of texture due to the internal degrees of freedom of $p$-wave
order parameters \cite{yip99,viljas99}.
In the above ways, order parameter configurations have been investigated
by using Josephson current-phase relations.

Since a wide variety of structures could be realized in systems with
internal degrees of freedom such as the textures in superfluid $^3$He
systems, we require a general theory for the Josephson effect to enable
us to identify the structures.
In this Letter, we derive a general formula for Josephson current,
applicable to a wide class of hybrid junctions between different
internal structures in unitary states, on the basis of the Andreev
picture \cite{andreev64} of the Josephson effect developed by Furusaki
and Tsukada (FT) \cite{furusaki91a}, including explicit expressions for
Andreev reflection coefficients reflecting their structures.

Consider a junction composed of three different superfluid regions
separated with flat interfaces perpendicular to the $z$-axis.
The interfaces are located at $z=0$ and $z=L$.
We shall denote these three regions by L ($z<0$), C ($0<z<L$), and
R ($L<z$), respectively.
We assume that the order parameter is uniform in each region.
We also assume that the effective mass $m$, Fermi velocity $v_F$, and
Fermi wave number $k_F$ are the same in all regions.
The potential barriers at the interfaces are ignored for simplicity.
We shall use a triad $(\hat{\bm{x}},\hat{\bm{y}},\hat{\bm{z}})$ as the
base for a frame of reference.

The dc Josephson current $I$ can be calculated from the temperature
Green's function, $\bm{\mathcal{G}}(z,z',\hat{\bm{k}}_{\|},\omega_n)$,
which is obtained by an analytic continuation,
$E+\text{i}0^+\rightarrow\text{i}\omega_n$, from the retarded Green's
function constructed by scattering data.
Here $\omega_n$ is the Matsubara frequency defined by
$\omega_n=\pi(2n+1)/\beta$ ($n=0,\pm1,\pm2,\ldots$) at temperature
$T=1/k_{\text{B}}\beta$ and
$\hat{\bm{k}}_{\|}\equiv(\hat{k}_x, \hat{k}_y,0)$ is defined from
$\hat{\bm{k}}=(\hat{k}_x, \hat{k}_y, \hat{k}_z)$ ($|\hat{\bm{k}}|=1$)
along the classical trajectory.
Note that the Green's functions are extended in the form of a $4\times4$
matrix due to the spin and particle-hole degrees of freedom required for
triplet pairing.
Taking the particle number conservation into account, the current at
$z=0$ is calculated by
\begin{equation}
 I=\frac{\hbar}{4\text{i}m\beta}\lim_{z,z'\rightarrow 0}\!
    \left(\frac{\partial}{\partial z}-\frac{\partial}{\partial z'}\right)
    \!\!\!\sum_{\omega_n,\hat{\bm{k}}_{\|}}\!\!\!
    \text{Tr}\,\bm{\mathcal{G}}(z,z',\hat{\bm{k}}_{\|},\omega_n).
    \label{eq:current}
\end{equation}

Let us calculate the scattering data necessary for constructing the
temperature Green's function.
In a quasiclassical approximation, the motion of a quasiparticle of
energy $E$ is described by Andreev equations applicable to an arbitrary
type of pairing \cite{yip85} as
\begin{equation}
 \begin{pmatrix}
  \left(-\text{i}\hbar v_F \hat{k}_z \frac{\partial}{\partial z}-E\right)
  \hat{\sigma}_0
  & \widehat{\Delta}^{\hat{\bm{k}}} \\
    \!\!\!\!\!\!\!\!\!\!\!\widehat{\Delta}^{\hat{\bm{k}}\dagger}
  & \!\!\!\!\!\!\!\!\!\!\!\left(\text{i}\hbar v_F \hat{k}_z
  \frac{\partial}{\partial z}-E\right) \hat{\sigma}_0
 \end{pmatrix}
 \begin{pmatrix}
  u \\ v
 \end{pmatrix}
 = 0,
\end{equation}
where
$u=(u_{\uparrow}\ u_{\downarrow})^t$ and
$v=(v_{\uparrow}\ v_{\downarrow})^t$.
$\hat{\sigma}_0$ is the $2\times 2$ unit matrix.
The gap matrix in spin space is denoted by
$\widehat{\Delta}^{\hat{\bm{k}}}$.
In each region
($\alpha=\text{L,C,R}$) the gap matrix,
$\widehat{\Delta}_{\alpha}^{\hat{\bm{k}}}$, is assumed to be constant
and unitary:
$\widehat{\Delta}_{\alpha}^{\hat{\bm{k}}}\cdot
\widehat{\Delta}_{\alpha}^{\hat{\bm{k}}\dagger}
=|\Delta_{\alpha}^{\hat{\bm{k}}}|^2 \hat{\sigma}_0$.
The solutions of the Andreev equations in each region are expressed as
\begin{align}
 \begin{pmatrix}
  u \\ v
 \end{pmatrix}
  & = \begin{pmatrix}
       (E+\Omega_{\alpha}^{\hat{\bm{k}}})\hat{\sigma}_0
       \text{e}^{\text{i}\nu_{\alpha} z}
       & \widehat{\Delta}_{\alpha}^{\hat{\bm{k}}}
       \text{e}^{-\text{i}\nu_{\alpha} z} \\
       \!\!\!\!\!\!\!\!\!\!\!\widehat{\Delta}_{\alpha}^{\hat{\bm{k}}\dagger}
       \text{e}^{\text{i}\nu_{\alpha} z}
       & \!\!\!\!\!\!\!\!\!\!\!(E+\Omega_{\alpha}^{\hat{\bm{k}}})
       \hat{\sigma}_0\text{e}^{-\text{i}\nu_{\alpha} z}
      \end{pmatrix}
  \begin{pmatrix}
   S_p \\ S_h 
  \end{pmatrix} \nonumber \\
 & \equiv \bm{M}_{\alpha}^{\hat{\bm{k}}}(z)\mathcal{S},
\end{align}
where 
$\Omega_{\alpha}^{\hat{\bm{k}}}
=\sqrt{E^2-|\Delta_{\alpha}^{\hat{\bm{k}}}|^2}$,
$S_p=(S_{p\uparrow}\ S_{p\downarrow})^t$, and
$S_h=(S_{h\uparrow}\ S_{h\downarrow})^t$.
$S_{p(h)\sigma}$ with $\sigma=\uparrow\ \text{or}\ \downarrow$ is the
probability amplitude for the mode propagating as a
particle-like (hole-like) quasiparticle with spin $\sigma$.
The two amplitudes $\mathcal{S}_{\text{L}}$ and $\mathcal{S}_{\text{R}}$
are connected by the continuity equation:
\begin{equation}
 \mathcal{S}_{\text{L}}
  = \bm{M}_{\text{L}}^{\hat{\bm{k}}}(0)^{-1}
  \bm{M}_{\text{C}}^{\hat{\bm{k}}}(0)
  \bm{M}_{\text{C}}^{\hat{\bm{k}}}(L)^{-1}
  \bm{M}_{\text{R}}^{\hat{\bm{k}}}(L)
  \mathcal{S}_{\text{R}}.
  \label{eq:continuity}
\end{equation}
Here we introduce scattering coefficient matrices in the following
form to express the scattering processes concisely;
\begin{equation}
 \hat{s}=\begin{pmatrix}
	  s_{\uparrow\uparrow} & s_{\uparrow\downarrow} \\
	  s_{\downarrow\uparrow} & s_{\downarrow\downarrow}
	 \end{pmatrix},
\end{equation}
where $s_{\sigma\sigma'}$ is the scattering coefficient for the process
in which an injected quasiparticle with spin $\sigma'$ is scattered as a
quasiparticle with spin $\sigma$.
The retarded Green's function is constructed from the scattering
coefficient matrices under the FT prescriptions \cite{furusaki91a}.
For $z\leq z' < 0$, the retarded Green's function is given by
\begin{align}
 & \bm{G}(z,z',\hat{\bm{k}}_{\|},E)
 = -\text{i}\frac{m}{\hbar^2k_F \hat{k}_z} \nonumber\\
 &\quad\times\left[
 \frac{\text{e}^{-\text{i}k_F \hat{k}_z(z-z')}
 \bm{M}_{\text{L}}^{\hat{\bm{k}}_-}(z)
 \genfrac{(}{)}{0pt}{1}{\,\hat{\sigma}_0 \,\  \hat{a}_2}{\hat{0} \ \ \ \hat{0}}
 \bm{M}_{\text{L}}^{\hat{\bm{k}}_-}(z')^{\dagger}
 }%
 {(E+\Omega_{\text{L}}^{\hat{\bm{k}}_-})^2
 -|\Delta_{\text{L}}^{\hat{\bm{k}}_-}|^2} \right.\nonumber\\
 &\quad\quad\quad + \left.
 \frac{\text{e}^{\text{i}k_F \hat{k}_z(z-z')}
 \bm{M}_{\text{L}}^{\hat{\bm{k}}}(z)
 \genfrac{(}{)}{0pt}{1}{\hat{0} \ \ \ \hat{0}}{\,\hat{a}_1 \,\ \hat{\sigma}_0}
 \bm{M}_{\text{L}}^{\hat{\bm{k}}}(z')^{\dagger}}%
 {(E+\Omega_{\text{L}}^{\hat{\bm{k}}})^2
 -|\Delta_{\text{L}}^{\hat{\bm{k}}}|^2}
 \right],
 \label{eq:green}
\end{align}
where $\hat{0}$ is the $2\times2$ zero matrix and
$\hat{\bm{k}}_-=\hat{\bm{k}}_{\|}-\hat{k}_z\hat{\bm{z}}$.
The Andreev reflection coefficients for a process in which a particle-like
quasiparticle injected from L is reflected as a hole-like quasiparticle
is obtained by solving Eq. (\ref{eq:continuity}) as
\begin{widetext}
 \begin{align}
  \hat{a}_1 & =
   \left[\text{e}^{-\text{i}\nu_{\text{c}} L}
     \left(\widehat{\Gamma}_{\text{C}}^{\hat{\bm{k}}\dagger}
     -\widehat{\Gamma}_{\text{L}}^{\hat{\bm{k}}\dagger}\right)\!
    \left(1-\widehat{\Gamma}_{\text{C}}^{\hat{\bm{k}}}\,
  \widehat{\Gamma}_{\text{R}}^{\hat{\bm{k}}\dagger}\right)
    +\text{e}^{\text{i}\nu_{\text{c}} L}
    \left(1-\widehat{\Gamma}_{\text{L}}^{\hat{\bm{k}}\dagger}\,
  \widehat{\Gamma}_{\text{C}}^{\hat{\bm{k}}}\right)\!
    \left(\widehat{\Gamma}_{\text{R}}^{\hat{\bm{k}}\dagger}
     -\widehat{\Gamma}_{\text{C}}^{\hat{\bm{k}}\dagger}\right)\right]
  \nonumber \\
  & \quad\quad\quad \times\left[\text{e}^{-\text{i}\nu_{\text{c}} L}
   \left(1-\widehat{\Gamma}_{\text{L}}^{\hat{\bm{k}}}\,
  \widehat{\Gamma}_{\text{C}}^{\hat{\bm{k}}\dagger}\right)\!
   \left(1-\widehat{\Gamma}_{\text{C}}^{\hat{\bm{k}}}\,
  \widehat{\Gamma}_{\text{R}}^{\hat{\bm{k}}\dagger}\right)
   +\text{e}^{\text{i}\nu_{\text{c}} L}
   \left(\widehat{\Gamma}_{\text{C}}^{\hat{\bm{k}}}
  -\widehat{\Gamma}_{\text{L}}^{\hat{\bm{k}}}\right)\!
   \left(\widehat{\Gamma}_{\text{R}}^{\hat{\bm{k}}\dagger}
    -\widehat{\Gamma}_{\text{C}}^{\hat{\bm{k}}\dagger}\right)\right]^{-1},
  \label{eq:andreev}
 \end{align}
\end{widetext}
where
$\widehat{\Gamma}_{\alpha}^{\hat{\bm{k}}}
=\widehat{\Delta}_{\alpha}^{\hat{\bm{k}}}/(E+\Omega_{\alpha}^{\hat{\bm{k}}})$
and 
$\nu_{\text{c}}=\Omega_{\text{C}}^{\hat{\bm{k}}}/\hbar v_F \hat{k}_z$.
The matrix $\hat{a}_2$ describes the reverse process.
The matrices $\hat{a}_2$ and $\hat{a}_1$ are related by
$\hat{a}_2(E,\hat{\bm{k}})=\hat{a}_1(E,\hat{\bm{k}}_-)^{\dagger}$.
Substituting the temperature Green's function obtained by the analytic
continuation into Eq.\ (\ref{eq:current}), a general formula for the
Josephson current at temperature $T$ is derived as
\begin{widetext}
  \begin{equation}
  I=-\frac{\,k_F^2}{4\hbar}\int\frac{\text{d}^2\hat{k}_{\|}}{(2\pi)^2}
   \frac{1}{\beta}
   \sum_{\omega_n}\text{Tr}
   \left[\frac{\widehat{\Delta}_{\text{L}}^{\hat{\bm{k}}}\,\hat{a}_1(\omega_n)
    +\hat{a}_1(\omega_n)\widehat{\Delta}_{\text{L}}^{\hat{\bm{k}}}}%
    {\widetilde{\Omega}_{n\text{L}}^{\hat{\bm{k}}}}
    -\frac{\hat{a}_2(\omega_n)
    \widehat{\Delta}_{\text{L}}^{\hat{\bm{k}}_-\dagger}
    +\widehat{\Delta}_{\text{L}}^{\hat{\bm{k}}_-\dagger}\hat{a}_2(\omega_n)}%
    {\widetilde{\Omega}_{n\text{L}}^{\hat{\bm{k}}_-}}\right],
   \label{eq:josephson}
  \end{equation}
 \end{widetext}
where $\widetilde{\Omega}_{n\text{L}}^{\hat{\bm{k}}}
=\sqrt{\omega_n^2+|\Delta_{\text{L}}^{\hat{\bm{k}}}|^2}$.
This is the central result of this Letter.
This formula is applicable to any type of Josephson junction between
unitary states with any symmetry.
Note that our formula still preserves the original FT form expressed by
the difference between Andreev reflection coefficients, describing the net
current carried by the two processes: the scattering of a particle-like
quasiparticle into a hole-like quasiparticle and its reverse process.
This formula can indeed be reduced to the FT formula by substituting the
$s$-wave order parameters.
Our formula is, therefore, a natural extension of the FT formula.

Let us consider certain special cases reported previously by several
authors.
Substituting the order parameters used in Ref.\ \onlinecite{yip99} into
our formulae (\ref{eq:andreev}) and (\ref{eq:josephson}), and adopting the
limit $L\rightarrow 0$, we obtained an analytic expression for the
Josephson current through a pinhole between two $^3$He-\textit{B}
reservoirs (Eq.\ (2) of Ref.\ \onlinecite{yip99}).
This reduces to the Kurkij\"arvi formula \cite{kurkijarvi88} for the
\textit{B} phase if the $\hat{\bm{n}}$ vectors of two \textit{B}-phase
reservoirs are parallel.
This is the same as the expression for the $s$-wave supercurrent through a
short orifice \cite{kulik77}.
By employing the \textit{A}-phase rather than the \textit{B}-phase order
parameter, the Kurkij\"arvi formula for the \textit{A} phase
\cite{kurkijarvi88} is reproduced.
Let us take another limit $\widehat{\Delta}_{\text{C}}\rightarrow 0$ with
a finite $L$.
This reproduces the current-phase relation of the
superconductor/normal-metal/superconductor (SNS) junctions without a
potential barrier \cite{furusaki91b}.
The Tanaka-Kashiwaya formula \cite{tanaka96} for unconventional
\textit{singlet} superconductors without a potential barrier is also
reproduced. 
In addition, the Yamashiro-Tanaka-Kashiwaya results \cite{yamashiro98} are
reproduced when we employ their order parameters.
Thus our formula covers the previous formulae for concrete examples.

\begin{figure}[tbp]
\includegraphics[keepaspectratio=true, width=0.46\textwidth]{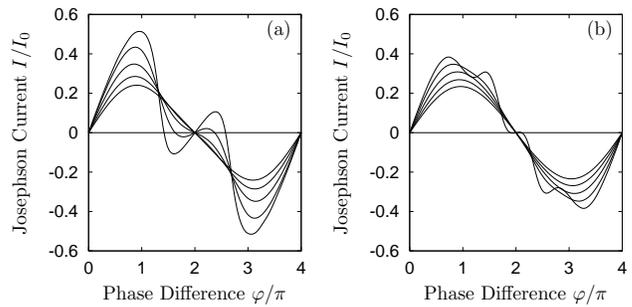}
\caption{\label{fig:Temp} Josephson current-phase relations of
 \textit{BAB} junctions for $L=2\xi_0$ at temperatures $0.1T_{\text{c}}$,
 $0.3T_{\text{c}}$, $0.5T_{\text{c}}$, $0.7T_{\text{c}}$, and
 $0.9T_{\text{c}}$ in decreasing order of gradient at the origin for the
 configurations of (a) \textbf{aa} and (b) \textbf{ab}. Here,
 $I_0=S\Delta_B(0)k_F^2/4\pi\hbar$ with $S$ being the interface area.}
\end{figure}

Furthermore, our formula enables us to analyze new kinds of junction.
One example is \textit{B}-phase/\textit{A}-phase/\textit{B}-phase
(\textit{BAB}) junctions in superfluid $^3$He systems.
This type of junction could be realized by using the experimental setup
used to study the Andreev reflection \cite{cousins96} at \textit{AB}
interfaces.
It is also suggested that a pseudo-\textit{A} phase could be formed
around the orifice in $^3$He-\textit{B} weak links \cite{packard00}.
In spin-\textit{triplet} Cooper-pair condensates, the boundary condition
should be carefully considered to include spin-orbit degrees of freedom.
The order parameter of the \textit{A} phase is defined by a triad
$(\hat{\bm{w}}_1,\hat{\bm{w}}_2,\hat{\bm{l}})$ in orbital space and a
vector $\hat{\bm{d}}$ in spin space.
For the \textit{B} phase we need a rotational matrix
$R(\hat{\bm{n}},\theta)$ with rotational axis $\hat{\bm{n}}$ and
rotational angle $\theta$ relating the spin space to the orbital space.
We take the phase angle for the \textit{A} phase to be zero.
For an \textit{AB} interface with dipole energy \cite{yip87,thuneberg92},
the boundary conditions are
 $\hat{\bm{l}}\parallel\hat{\bm{d}}\perp\hat{\bm{z}}$,
 $\hat{\bm{d}}=R(\hat{\bm{n}},\theta_L)\hat{\bm{w}}_1$, and
 $\hat{\bm{w}}_1\parallel\hat{\bm{z}}$,
where $\theta_L=\cos^{-1}(-1/4)$ (Leggett angle).
Since the \text{A} phase always has the ambiguity of
$\hat{\bm{d}}\rightarrow-\hat{\bm{d}}$,
$\hat{\bm{w}}_1\rightarrow-\hat{\bm{w}}_1$, and
$\hat{\bm{w}}_2\rightarrow-\hat{\bm{w}}_2$,
we choose $\hat{\bm{w}}_1=\hat{\bm{z}}$, $\hat{\bm{w}}_2=\hat{\bm{x}}$,
and $\hat{\bm{l}}=\hat{\bm{y}}$ without loss of generality.
The vector $\hat{\bm{n}}$ in the \textit{B} phase is fixed in one of four
directions, \textit{i.e.},
$\left(-\sqrt{3/5}, \pm\sqrt{1/5}, \pm\sqrt{1/5}\right)$ for
$\hat{\bm{d}}=\hat{\bm{l}}$ and
$\left(\sqrt{3/5}, \mp\sqrt{1/5}, \pm\sqrt{1/5}\right)$ for
$\hat{\bm{d}}=-\hat{\bm{l}}$. 
We use the letters \textbf{a}, \textbf{b}, \textbf{c}, and
\textbf{d} in turn to specify the different orientations of
$\hat{\bm{n}}$ for brevity \cite{yip99}.
The order pairs of the letters \textbf{aa}, \textbf{ab}, etc., thus
denote the order parameter configurations at the two interfaces.
The order parameters for the configurations of \textbf{aa} and
\textbf{ab}, which become important in the following discussions, are
given by
\begin{align}
 &\widehat{\Delta}_{\text{L}}^{\hat{\bm{k}}}
 =\Delta_{B}\text{e}^{-\text{i}\varphi/2}/2\nonumber\\
 &\ \times\left\{2\text{i}\hat{\sigma}_0\hat{k}_z
 +\hat{\sigma}_1(\sqrt{3}k_x-k_y)
 -\hat{\sigma}_3(k_x+\sqrt{3}k_y)\right\}
 \nonumber\\
 &\widehat{\Delta}_{\text{C}}^{\hat{\bm{k}}}
 =\Delta_{A}(k_x-\text{i}\hat{k}_z)\hat{\sigma}_0 \nonumber\\
 &\widehat{\Delta}_{\text{R}}^{\hat{\bm{k}}}
 =\Delta_{B}\text{e}^{\text{i}\varphi/2}/2\nonumber\\
 &\ \times\left\{2\text{i}\hat{\sigma}_0\hat{k}_z
 +\hat{\sigma}_1(\pm\sqrt{3}k_x-k_y)
 -\hat{\sigma}_3(k_x\pm\sqrt{3}k_y)\right\},
\end{align}
where $\varphi$ is the phase difference and
$(\hat{\sigma}_1,\hat{\sigma}_2,\hat{\sigma}_3)$ are Pauli's matrices.
$\Delta_{A}$ and $\Delta_{B}$ are the gaps for the \textit{A} phase and
the \textit{B} phase, respectively.
A rotation of $\pi$ around the $y$-axis affects the transformations
\textbf{a}$\leftrightarrow$\textbf{d},
\textbf{b}$\leftrightarrow$\textbf{c},
L$\leftrightarrow$R, and
$\hat{\bm{d}}\rightarrow-\hat{\bm{d}}$, where we used the ambiguity of
the \textit{A} phase.
In addition, \textbf{aa} and \textbf{bb} configurations give the
identical current, same as \textbf{ab} and \textbf{ba}.
As a result, we predict two types of current-phase characteristics:
type (a) \{\textbf{aa},\textbf{bb},\textbf{cc},\textbf{dd}\} and
type (b) \{\textbf{ab},\textbf{ba},\textbf{cd},\textbf{dc}\}.
Figure \ref{fig:Temp} shows the temperature dependence of the Josephson
current in (a) \textbf{aa} and (b) \textbf{ab} configurations for
$L=2\xi_0$ where  the coherence length $\xi_0$ in the \textit{B} phase at
zero temperature is given by $\xi_0=[7\zeta(3)/48]^{1/2}(\hbar v_F/\pi
k_{\text{B}}T_{\text{c}})$ with $\zeta$ being the Riemann zeta-function.
Here we assume the gaps obey the BCS temperature dependence.
We can observe current-phase curves with $4\pi$ rather than $2\pi$
periodicity and find some peak structures at lower temperatures.
In particular, the structures become pronounced as the temperature
decreases.
In the Andreev picture, the Josephson current is carried by both a
discrete part (Andreev bound states) and a continuous part of the
excitation spectrum.
This suggests that the peak structures and the $4\pi$ periodicity come
from the Andreev bound states and the continuous spectrum, respectively.
\begin{figure}[tbp]
\includegraphics[keepaspectratio=true, width=0.46\textwidth]{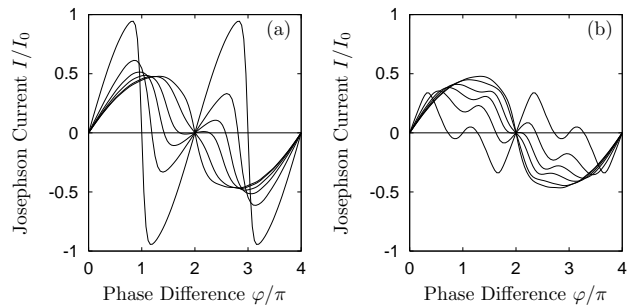}
\caption{\label{fig:Width} Josephson current-phase relations of
 \textit{BAB} junctions at $0.1T_{\text{c}}$ for $L=0$, $\xi_0$,
 $2\xi_0$, $3\xi_0$, $5\xi_0$, and $1000\xi_0$, in decreasing order of
 gradient at the origin for the configurations of (a) \textbf{aa} and
 (b) \textbf{ab}.}
\end{figure}

It is known that Andreev bound states are localized near an interface.
In the system with two interfaces that we consider here, the interaction
between bound states near the two interfaces results in the current
depending on the distance between the interfaces \cite{hurd95,nakayama00}.
In order to clarify the bound state contribution to the current, we
investigate the current-phase relations of junctions with different
\textit{A}-phase widths at low temperatures.
Figure \ref{fig:Width} clearly shows that the peak structures are
pronounced, while the $4\pi$ component is suppressed as $L$ decreases.
When $L\gg\xi_0$, the Andreev bound states near the two interfaces are
isolated and their overlap is lost.
A \textit{BAB} junction is regarded as a series of \textit{BA} and
\textit{AB} junctions.
Since the total phase difference through the \textit{BAB} junction is
$\varphi$, the \textit{BA} (\textit{AB}) junction has $\varphi/2$
dependence, resulting the $4\pi$ periodicity in the current-phase
characteristics.
In addition, the current-phase relations in (a) and (b) merge into
a single curve in the limit $L\rightarrow\infty$ because the current
contributions from isolated interfaces are the same.
When $L\rightarrow0$, our model is reduced to the pinhole model in
$^3$He-\textit{B} under a sufficiently large magnetic field applied
parallel to the wall \cite{yip99}.
The surface free energy $-(\bm{H}\cdot R\hat{\bm{z}})^2$ produced by
the applied magnetic field $\bm{H}$ is minimized if $\hat{\bm{n}}$
aligns in one of the four directions, \textbf{a}, \textbf{b},
\textbf{c}, or \textbf{d}.
Therefore, $\hat{\bm{l}}$ plays a similar role to $\bm{H}$ in
determining $\hat{\bm{n}}$ directions.
This could be regarded as being a ``pseudo-magnetic interface effect''
that is inherent in \textit{BAB} junction systems.
The difference between $\bm{H}$ and the \textit{AB} interface appears in
the number of allowed $\hat{\bm{n}}$ directions: all four directions
for $\bm{H}$, while only two directions in each
$\hat{\bm{d}}=\pm\hat{\bm{l}}$ for the \textit{AB} interfaces.
Thus the curve for $L=0$ in Fig.\ \ref{fig:Width}(b) is simply Yip's
$\pi$ state  \cite{yip99}, that is produced by the current cancellation
due to quasiparticles with different $\hat{\bm{n}}$-dependent effective
phases.
The curve in Fig.\ \ref{fig:Width}(a) corresponds to the curve for
``AA'' in Fig.\ 4 of Ref.\ \onlinecite{yip99}.
From the above discussion, the curve for ``AC'' cannot exist in a
\textit{BAB} junction.
For $L\sim\xi_0$, the Andreev bound states near two interfaces maintain
their overlap and the \textit{BAB} junction behaves like a single
junction.
The current-phase relation still maintains the characteristics of the
$L=0$ curve.

Now let us revisit Fig.\ \ref{fig:Temp}.
The bound states localized near each interface overlap because
$L\sim\xi_0$.
Since the current contribution due to the Andreev bound states is
dominant at low temperatures, $2\pi$-periodic current-phase
characteristics appear in the curves as peak structures.
As the temperature becomes higher, the continuous spectrum contribution
becomes large and obscures the Andreev bound state contribution.
Therefore, we conclude that the peak structures appearing in Fig.\
\ref{fig:Temp} come from the Andreev bound states and the $4\pi$
periodicity is brought about by the continuous spectrum.

Throughout the above studies of \textit{BAB} junctions, the
current-phase characteristics reflected the configurations of textures
and were modified by arranging the textures of each phase.
Since our formula applies to any type of Josephson junction, systematic
studies of current-phase relations reveal the internal structures due to
spin-orbit degrees of freedom as well as their pairing symmetries.

In summary, we derived a general formula for Josephson current based
on the Andreev picture.
This formula made it possible to analyze any type of Josephson
junction between unitary states with any symmetry, and was applied
to novel \textit{BAB} junctions in superfluid $^3$He, which are
accessible to experiments.
It has been predicted that there are two types of current-phase
relations associated with \textit{AB} interfaces inherent in the system,
acting similarly to magnetic field $\bm{H}$ through $\hat{\bm{l}}$ in
the \textit{A} phase (``pseudo-magnetic interface effect'').

\begin{acknowledgments}
 We thank E. V. Thuneberg, J. Viljas, G. E. Volovik, M. Krusius,
 L. Skrv\'ek, J. Kopu, and D. Meacock for stimulating discussions.
 We are grateful to Y. Tanaka and Y. Ohashi for valuable input.
 We also thank Y. Asano for sending us his preprint.
 N. H. is indebted to H. Takayanagi and M. Paalanen for their continuous
 encouragement.
 This work was supported in part by the Waseda University Grant for
 Special Research Projects (2000A-875), by the COE Program ``Molecular
 Nano-Engineering'' from the Ministry of Education, Science and Culture,
 Japan, and by the Helsinki University of Technology and the National
 Technology Agency Tekes. 
\end{acknowledgments}
\textit{Note added.}---
During the preparation of this Letter, we learned of a theory
\cite{asano01} related to Josephson current in a similar system.
%\bibliography{bab}% Produces the bibliography via BibTeX.

\begin{thebibliography}{25}
\expandafter\ifx\csname natexlab\endcsname\relax\def\natexlab#1{#1}\fi
\expandafter\ifx\csname bibnamefont\endcsname\relax
  \def\bibnamefont#1{#1}\fi
\expandafter\ifx\csname bibfnamefont\endcsname\relax
  \def\bibfnamefont#1{#1}\fi
\expandafter\ifx\csname citenamefont\endcsname\relax
  \def\citenamefont#1{#1}\fi
\expandafter\ifx\csname url\endcsname\relax
  \def\url#1{\texttt{#1}}\fi
\expandafter\ifx\csname urlprefix\endcsname\relax\def\urlprefix{URL }\fi
\providecommand{\bibinfo}[2]{#2}
\providecommand{\eprint}[2][]{\url{#2}}

\bibitem[{\citenamefont{Josephson}(1965)}]{josephson65}
\bibinfo{author}{\bibfnamefont{B.~D.} \bibnamefont{Josephson}},
  \bibinfo{journal}{Adv.\ Phys.} \textbf{\bibinfo{volume}{14}},
  \bibinfo{pages}{419} (\bibinfo{year}{1965}).

\bibitem[{\citenamefont{Sigrist and Rice}(1992)}]{sigrist92}
\bibinfo{author}{\bibfnamefont{M.}~\bibnamefont{Sigrist}} \bibnamefont{and}
  \bibinfo{author}{\bibfnamefont{T.~M.} \bibnamefont{Rice}},
  \bibinfo{journal}{J. Phys.\ Soc.\ Jpn.} \textbf{\bibinfo{volume}{61}},
  \bibinfo{pages}{4283} (\bibinfo{year}{1992}).

\bibitem[{\citenamefont{Wollman et~al.}(1993)\citenamefont{Wollman, Harlingen,
  Lee, Ginsberg, and Leggett}}]{wollman93}
\bibinfo{author}{\bibfnamefont{D.~A.} \bibnamefont{Wollman}} \textit{et al}.,
%  \bibinfo{author}{\bibfnamefont{D.~J.~V.} \bibnamefont{Harlingen}},
%  \bibinfo{author}{\bibfnamefont{W.~C.} \bibnamefont{Lee}},
%  \bibinfo{author}{\bibfnamefont{D.~M.} \bibnamefont{Ginsberg}},
%  \bibnamefont{and} \bibinfo{author}{\bibfnamefont{A.~J.}
%  \bibnamefont{Leggett}},
  \bibinfo{journal}{Phys.\ Rev.\ Lett.}
  \textbf{\bibinfo{volume}{71}}, \bibinfo{pages}{2134} (\bibinfo{year}{1993}).

\bibitem[{\citenamefont{Jin et~al.}(1999)\citenamefont{Jin, Zadorozhny, Schlom,
  Mori, Maeno, and Liu}}]{jin99}
\bibinfo{author}{\bibfnamefont{R.}~\bibnamefont{Jin}} \textit{et al}.,
%  \bibinfo{author}{\bibfnamefont{Y.}~\bibnamefont{Zadorozhny}},
%  \bibinfo{author}{\bibfnamefont{D.~G.} \bibnamefont{Schlom}},
%  \bibinfo{author}{\bibfnamefont{Y.}~\bibnamefont{Mori}},
%  \bibinfo{author}{\bibfnamefont{Y.}~\bibnamefont{Maeno}}, \bibnamefont{and}
%  \bibinfo{author}{\bibfnamefont{Y.}~\bibnamefont{Liu}},
  \bibinfo{journal}{Phys.\ Rev.\ B} \textbf{\bibinfo{volume}{59}},
  \bibinfo{pages}{4433} (\bibinfo{year}{1999}).

\bibitem[{\citenamefont{Yamashiro et~al.}(1998)\citenamefont{Yamashiro, Tanaka,
  and Kashiwaya}}]{yamashiro98}
\bibinfo{author}{\bibfnamefont{M.}~\bibnamefont{Yamashiro}},
  \bibinfo{author}{\bibfnamefont{Y.}~\bibnamefont{Tanaka}}, \bibnamefont{and}
  \bibinfo{author}{\bibfnamefont{S.}~\bibnamefont{Kashiwaya}},
  \bibinfo{journal}{J. Phys.\ Soc.\ Jpn.} \textbf{\bibinfo{volume}{67}},
  \bibinfo{pages}{3364} (\bibinfo{year}{1998}).

\bibitem[{\citenamefont{Honerkamp and Sigrist}(1998)}]{honerkamp98}
\bibinfo{author}{\bibfnamefont{C.}~\bibnamefont{Honerkamp}} \bibnamefont{and}
  \bibinfo{author}{\bibfnamefont{M.}~\bibnamefont{Sigrist}},
  \bibinfo{journal}{Prog.\ Theor.\ Phys.} \textbf{\bibinfo{volume}{100}},
  \bibinfo{pages}{53} (\bibinfo{year}{1998}).

\bibitem[{\citenamefont{Backhaus et~al.}(1998)\citenamefont{Backhaus,
  Pereverzev, Simmonds, Loshak, Davis, and Packard}}]{packard98}
\bibinfo{author}{\bibfnamefont{S.}~\bibnamefont{Backhaus}} \textit{et al}.,
%  \bibinfo{author}{\bibfnamefont{S.~V.} \bibnamefont{Pereverzev}},
%  \bibinfo{author}{\bibfnamefont{R.~W.} \bibnamefont{Simmonds}},
%  \bibinfo{author}{\bibfnamefont{A.}~\bibnamefont{Loshak}},
%  \bibinfo{author}{\bibfnamefont{J.~C.} \bibnamefont{Davis}}, \bibnamefont{and}
%  \bibinfo{author}{\bibfnamefont{R.~E.} \bibnamefont{Packard}},
  \bibinfo{journal}{Nature} \textbf{\bibinfo{volume}{392}},
  \bibinfo{pages}{687} (\bibinfo{year}{1998}).

\bibitem[{\citenamefont{Marchenkov et~al.}(1999)\citenamefont{Marchenkov,
  Simmonds, Backhaus, Loshak, Davis, and Packard}}]{packard99}
\bibinfo{author}{\bibfnamefont{A.}~\bibnamefont{Marchenkov}} \textit{et al}.,
%  \bibinfo{author}{\bibfnamefont{R.~W.} \bibnamefont{Simmonds}},
%  \bibinfo{author}{\bibfnamefont{S.}~\bibnamefont{Backhaus}},
%  \bibinfo{author}{\bibfnamefont{A.}~\bibnamefont{Loshak}},
%  \bibinfo{author}{\bibfnamefont{J.~C.} \bibnamefont{Davis}}, \bibnamefont{and}
%  \bibinfo{author}{\bibfnamefont{R.~E.} \bibnamefont{Packard}},
  \bibinfo{journal}{Phys.\ Rev.\ Lett.} \textbf{\bibinfo{volume}{83}},
  \bibinfo{pages}{3860} (\bibinfo{year}{1999}).

\bibitem[{\citenamefont{Avenel et~al.}(2000)\citenamefont{Avenel, Mukharsky,
  and Varoquaux}}]{avenel00}
\bibinfo{author}{\bibfnamefont{O.}~\bibnamefont{Avenel}},
  \bibinfo{author}{\bibfnamefont{Y.}~\bibnamefont{Mukharsky}},
  \bibnamefont{and}
  \bibinfo{author}{\bibfnamefont{E.}~\bibnamefont{Varoquaux}},
  \bibinfo{journal}{Physica B} \textbf{\bibinfo{volume}{280}},
  \bibinfo{pages}{130} (\bibinfo{year}{2000}).

\bibitem[{\citenamefont{Yip}(1999)}]{yip99}
\bibinfo{author}{\bibfnamefont{S.-K.} \bibnamefont{Yip}},
  \bibinfo{journal}{Phys.\ Rev.\ Lett.} \textbf{\bibinfo{volume}{83}},
  \bibinfo{pages}{3864} (\bibinfo{year}{1999}).

\bibitem[{\citenamefont{Viljas and Thuneberg}(1999)}]{viljas99}
\bibinfo{author}{\bibfnamefont{J.~K.} \bibnamefont{Viljas}} \bibnamefont{and}
  \bibinfo{author}{\bibfnamefont{E.~V.} \bibnamefont{Thuneberg}},
  \bibinfo{journal}{Phys.\ Rev.\ Lett.} \textbf{\bibinfo{volume}{83}},
  \bibinfo{pages}{3868} (\bibinfo{year}{1999}).

\bibitem[{\citenamefont{Andreev}(1964)}]{andreev64}
\bibinfo{author}{\bibfnamefont{A.~F.} \bibnamefont{Andreev}},
  \bibinfo{journal}{Zh.\ Eksp.\ Teor.\ Fiz.} \textbf{\bibinfo{volume}{46}},
  \bibinfo{pages}{1823} (\bibinfo{year}{1964}) [Sov.\ Phys.\
  JETP \textbf{19}, 1228 (1964)].

\bibitem[{\citenamefont{Furusaki and
  Tsukada}(1991{\natexlab{a}})}]{furusaki91a}
\bibinfo{author}{\bibfnamefont{A.}~\bibnamefont{Furusaki}} \bibnamefont{and}
  \bibinfo{author}{\bibfnamefont{M.}~\bibnamefont{Tsukada}},
  \bibinfo{journal}{Solid State Commun.} \textbf{\bibinfo{volume}{78}},
  \bibinfo{pages}{299} (\bibinfo{year}{1991}{\natexlab{a}}).

\bibitem[{\citenamefont{Yip}(1985)}]{yip85}
\bibinfo{author}{\bibfnamefont{S.}~\bibnamefont{Yip}}, \bibinfo{journal}{Phys.\
  Rev.\ B} \textbf{\bibinfo{volume}{32}}, \bibinfo{pages}{2915}
  (\bibinfo{year}{1985}).

\bibitem[{\citenamefont{Kurkij\"arvi}(1988)}]{kurkijarvi88}
\bibinfo{author}{\bibfnamefont{J.}~\bibnamefont{Kurkij\"arvi}},
  \bibinfo{journal}{Phys.\ Rev.\ B} \textbf{\bibinfo{volume}{38}},
  \bibinfo{pages}{11184} (\bibinfo{year}{1988}).

\bibitem[{\citenamefont{Kulik and Omel'yanchuk}(1977)}]{kulik77}
\bibinfo{author}{\bibfnamefont{I.~O.} \bibnamefont{Kulik}} \bibnamefont{and}
  \bibinfo{author}{\bibfnamefont{A.~N.} \bibnamefont{Omel'yanchuk}},
  \bibinfo{journal}{Fiz. Nizk. Temp.} \textbf{\bibinfo{volume}{3}},
  \bibinfo{pages}{945} (\bibinfo{year}{1977}) [Sov. J. Low
  Temp.\ Phys.\ \textbf{3}, 459 (1977)].

\bibitem[{\citenamefont{Furusaki and
  Tsukada}(1991{\natexlab{b}})}]{furusaki91b}
\bibinfo{author}{\bibfnamefont{A.}~\bibnamefont{Furusaki}} \bibnamefont{and}
  \bibinfo{author}{\bibfnamefont{M.}~\bibnamefont{Tsukada}},
  \bibinfo{journal}{Phys.\ Rev.\ B} \textbf{\bibinfo{volume}{43}},
  \bibinfo{pages}{10164} (\bibinfo{year}{1991}{\natexlab{b}}).

\bibitem[{\citenamefont{Tanaka and Kashiwaya}(1996)}]{tanaka96}
\bibinfo{author}{\bibfnamefont{Y.}~\bibnamefont{Tanaka}} \bibnamefont{and}
  \bibinfo{author}{\bibfnamefont{S.}~\bibnamefont{Kashiwaya}},
  \bibinfo{journal}{Phys.\ Rev.\ B} \textbf{\bibinfo{volume}{53}},
  \bibinfo{pages}{R11957} (\bibinfo{year}{1996}).

\bibitem[{\citenamefont{Cousins et~al.}(1996)\citenamefont{Cousins, Enrico,
  Fisher, Phillipson, Pickett, Shaw, and Thibault}}]{cousins96}
\bibinfo{author}{\bibfnamefont{D.~J.} \bibnamefont{Cousins}} \textit{et al}.,
%  \bibinfo{author}{\bibfnamefont{M.~P.} \bibnamefont{Enrico}},
%  \bibinfo{author}{\bibfnamefont{S.~N.} \bibnamefont{Fisher}},
%  \bibinfo{author}{\bibfnamefont{S.~L.} \bibnamefont{Phillipson}},
%  \bibinfo{author}{\bibfnamefont{G.~R.} \bibnamefont{Pickett}},
%  \bibinfo{author}{\bibfnamefont{N.~S.} \bibnamefont{Shaw}}, \bibnamefont{and}
%  \bibinfo{author}{\bibfnamefont{P.~J.~Y.} \bibnamefont{Thibault}},
  \bibinfo{journal}{Phys.\ Rev.\ Lett.} \textbf{\bibinfo{volume}{77}},
  \bibinfo{pages}{5245} (\bibinfo{year}{1996}).

\bibitem[{\citenamefont{Simmonds et~al.}(2000)\citenamefont{Simmonds,
  Marchenkov, Vitale, Davis, and Packard}}]{packard00}
\bibinfo{author}{\bibfnamefont{R.~W.} \bibnamefont{Simmonds}}
  \textit{et al}.,
%  \bibinfo{author}{\bibfnamefont{A.}~\bibnamefont{Marchenkov}},
%  \bibinfo{author}{\bibfnamefont{S.}~\bibnamefont{Vitale}},
%  \bibinfo{author}{\bibfnamefont{J.~C.} \bibnamefont{Davis}}, \bibnamefont{and}
%  \bibinfo{author}{\bibfnamefont{R.~E.} \bibnamefont{Packard}},
  \bibinfo{journal}{Phys.\ Rev.\ Lett.} \textbf{\bibinfo{volume}{84}},
  \bibinfo{pages}{6062} (\bibinfo{year}{2000}).

\bibitem[{\citenamefont{Yip}(1987)}]{yip87}
\bibinfo{author}{\bibfnamefont{S.}~\bibnamefont{Yip}}, \bibinfo{journal}{Phys.\
  Rev.\ B} \textbf{\bibinfo{volume}{35}}, \bibinfo{pages}{8733}
  (\bibinfo{year}{1987}).

\bibitem[{\citenamefont{Thuneberg}(1992)}]{thuneberg92}
\bibinfo{author}{\bibfnamefont{E.~V.} \bibnamefont{Thuneberg}},
  \bibinfo{journal}{Physica B} \textbf{\bibinfo{volume}{178}},
  \bibinfo{pages}{168} (\bibinfo{year}{1992}).

\bibitem[{\citenamefont{Hurd and Wendin}(1995)}]{hurd95}
\bibinfo{author}{\bibfnamefont{M.}~\bibnamefont{Hurd}} \bibnamefont{and}
  \bibinfo{author}{\bibfnamefont{G.}~\bibnamefont{Wendin}},
  \bibinfo{journal}{Phys.\ Rev.\ B} \textbf{\bibinfo{volume}{51}},
  \bibinfo{pages}{3754} (\bibinfo{year}{1995}).

\bibitem[{\citenamefont{Nakayama et~al.}(2000)\citenamefont{Nakayama, Furukawa,
  and Okabe}}]{nakayama00}
\bibinfo{author}{\bibfnamefont{A.}~\bibnamefont{Nakayama}},
  \bibinfo{author}{\bibfnamefont{T.}~\bibnamefont{Furukawa}}, \bibnamefont{and}
  \bibinfo{author}{\bibfnamefont{Y.}~\bibnamefont{Okabe}}, \bibinfo{journal}{J.
  Appl.\ Phys.} \textbf{\bibinfo{volume}{88}}, \bibinfo{pages}{6605}
  (\bibinfo{year}{2000}).

\bibitem[{\citenamefont{Asano}()}]{asano01}
\bibinfo{author}{\bibfnamefont{Y.}~\bibnamefont{Asano}},
  \bibinfo{note}{preprint}.

\end{thebibliography}

%
%\begin{figure}[p]
%\includegraphics{fig1.eps}
%\caption{\label{fig:Temp} Josephson current-phase relations of
% \textit{BAB} junctions for $L=2\xi_0$ at temperatures $0.1T_{\text{c}}$,
% $0.3T_{\text{c}}$, $0.5T_{\text{c}}$, $0.7T_{\text{c}}$, and
% $0.9T_{\text{c}}$ in decreasing order of gradient at the origin for the
% configurations of (a) \textbf{aa} and (b) \textbf{ab}. Here,
% $I_0=S\Delta_B(0)k_F^2/4\pi\hbar$ with $S$ being the interface area.}
%\end{figure}
%
%\begin{figure}[p]
%\includegraphics{fig2.eps}
%\caption{\label{fig:Width} Josephson current-phase relations of
% \textit{BAB} junctions at $0.1T_{\text{c}}$ for $L=0$, $\xi_0$,
% $2\xi_0$, $3\xi_0$, $5\xi_0$, and $1000\xi_0$, in decreasing order of
% gradient at the origin for the configurations of (a) \textbf{aa} and
% (b) \textbf{ab}.}
%\end{figure}
\end{document}